\documentclass[reprint,aps,prb,amsmath,amssymb,twocolumn,showkeys, superscriptaddress,floatfix,showkeys]{revtex4-2}
\usepackage{graphicx}
\usepackage{dcolumn}
\usepackage{bm}
\usepackage[colorlinks = true, allcolors = blue]{hyperref}
\usepackage{verbatim}
\usepackage{soul}
\usepackage{float}
\usepackage{xcolor}
\usepackage{multirow}

\begin{document}

\title{Retrieval of fundamental material parameters of monolayer transition metal dichalcogenides from experimental exciton energies: An analytical approach}

\author{Duy-Nhat Ly}
\email{nhatld@hcmue.edu.vn}
\affiliation{Computational Physics Key Laboratory K002, Department of Physics, Ho Chi Minh City University of Education, Ho Chi Minh City 72759, Vietnam}

\author{Dai-Nam Le}
\affiliation{Department of Physics, University of South Florida, Tampa, Florida 33620, USA}

\author{Dang-Khoa D. Le}
\affiliation{Computational Physics Key Laboratory K002, Department of Physics, Ho Chi Minh City University of Education, Ho Chi Minh City 72759, Vietnam}

\author{Van-Hoang Le}
\email{hoanglv@hcmue.edu.vn}
\affiliation{Computational Physics Key Laboratory K002, Department of Physics, Ho Chi Minh City University of Education, Ho Chi Minh City 72759, Vietnam}

\date{\today}

\begin{abstract}
We propose a straightforward and highly accurate method for extracting material parameters such as screening length, bandgap energy, exciton reduced mass, and the dielectric constant of the surrounding medium from experimental magnetoexciton energies available for monolayer transition metal dichalcogenides (TMDCs). Our approach relies on analytical formulations that allow us to calculate the screening length $r_0$ and bandgap energy $E_g$ directly from the experimental $s$-state exciton energies $E_{1s}$, $E_{2s}$, and $E_{3s}$. We also establish a relationship between the surrounding dielectric constant $\kappa$ and the exciton reduced mass $\mu$. This relationship simplifies the Schr{\"o}dinger equation for a magnetoexciton in a TMDC monolayer, transforming it into a one-parameter equation that depends solely on the single material parameter $\mu$.
Furthermore, we develop an analytical formula with high accuracy for magnetoexciton energies as a function of the exciton reduced mass: $E(B,\mu)$. Then, the inverse of this formula allows us to calculate the exciton reduced mass from experimental data on magnetoexciton energies. By applying this method, we extract key material parameters, $E_g$, $r_0$, $\mu$, and $\kappa$, from the magnetoexciton energies of monolayer TMDCs, including WSe$_2$, WS$_2$, MoSe$_2$, and MoS$_2$, encapsulated by hexagonal boron nitride (hBN) slabs in various current experiments. The material properties we retrieve complement and correct existing experimental and theoretical data. Additionally, we develop an analytical method for calculating diamagnetic coefficients and exciton radii with high accuracy compared to numerical calculations. Based on this method, we provide diamagnetic coefficients and exciton radii computed using the extracted material parameters.

\end{abstract}


\maketitle

\section{Introduction}
 Two-dimensional semiconductors with thickness of atomic scales, such as monolayer transition metal dichalcogenides (TMDCs), have attracted significant attention over the past decade because they enable the fabrication of ultra-thin electronic devices with exceptional performance \cite{ Liu2018CSV, Liu2021pers, Das2021, Huang2022}. 
High promises of these materials come from their incredible properties, which include bandgap tunability, high optical sensitivity, strong spin-orbit coupling, valley polarization, considerable excitonic binding energy, and the high thermal stability of excitons at room temperature \cite{HePRL2014, Ugeda2014, Ye2014, Arora2015, Hill2015, Wang2018}. Therefore, a precise understanding of the intrinsic optoelectric properties of these TMDCs, such as their energy bandgap $E_g$, exciton reduced mass $\mu$, screening length $r_0$ (related to the two-dimensional (2D) static polarizability $\chi_{2D}$), and the effective dielectric constant $\kappa$ of the surrounding medium, is essential for advancing the development of TMDC–based devices. 

Several methods are available for determining the optoelectric properties of TMDCs. For instance, the exciton reduced mass can be directly determined by extracting effective electron and hole masses from angle-resolved photoemission spectroscopy (ARPES), which allows for the experimental detection of energy versus momentum maps \cite{Basov2014, Bussolotti2021, Lee2021, Lin2022}. Meanwhile, the bandgap energy can be probed by time-resolved ARPES \cite{Lee2021}, scanning tunneling microscopy (STM) \cite{Zhang2014nanoletters, Huang2015natcom, Trainer2017}, or photoluminescence excitation (PLE) spectra \cite{Zhu2015, Yao2017}. From a theoretical perspective, \textit{ab initio} calculations based on density functional theory (DFT), time-dependent DFT, or the GW-Bethe-Salpeter equation (GW-BSE) can provide predictions for the exciton reduced mass, bandgap energy, or even 2D polarizability \cite{xiao2012, berkelbach2013, kylanpaa2015, Korm2015, haastrup2018, gjerding2021, Gulevich2023}. Unfortunately, balancing accuracy and cost among these experimental measurements or theoretical calculations is challenging. 

Nevertheless, exciton spectroscopy offers alternative methods to probe the intrinsic optoelectric properties of materials, combining theoretical knowledge of exciton spectra with experimental measurements via optical spectroscopies. Recent studies \cite{Stier2016-nano, Stier2018, NAT2019, Liu2019} have revealed the exciton reduced mass $\mu$ from optical spectroscopy of magnetoexcitons at high magnetic fields (65 to 91 tesla), like that people did for 3D perovskite semiconductors \cite{Miyata2015, Reimann1998}. However, such magnetic fields are not theoretically large enough for their approximation (see a discussion in Ref.~ \cite{Nhat2023May}). Another promising approach is to utilize both PLE and two-photon measurements, as has been done in 3D semiconductors \cite{Reimann1998}. This approach requires a theoretical model to describe the relationship between the exciton spectra of $s$ and $p$ states, which is currently unavailable due to the nonhydrogenic nature of exciton spectra in TMDC monolayers. Indeed, in this case, unlike 3D semiconductors, the dimensionality reduction causes a screening effect in the Coulomb interaction between electron and hole \cite{berkelbach2013, chernikov2014, ChernikovLett2015}, which makes them coupled by the Rytova-Keldysh potential \cite{keldysh1979}. Thus, exciton spectra in TMDC monolayers are much more complicated than the well-known 2D hydrogenic model \cite{Olsen2016, Molas2019, Chennano2019, takahashi2024, Hsu2DMater2019, Liu2021,Hieu2022, Kezerashvili2023, Pedersen2023, Doma2023, Dinh2025}.  

\begin{figure}[H]
\begin{center}
\includegraphics[width=1. \columnwidth]{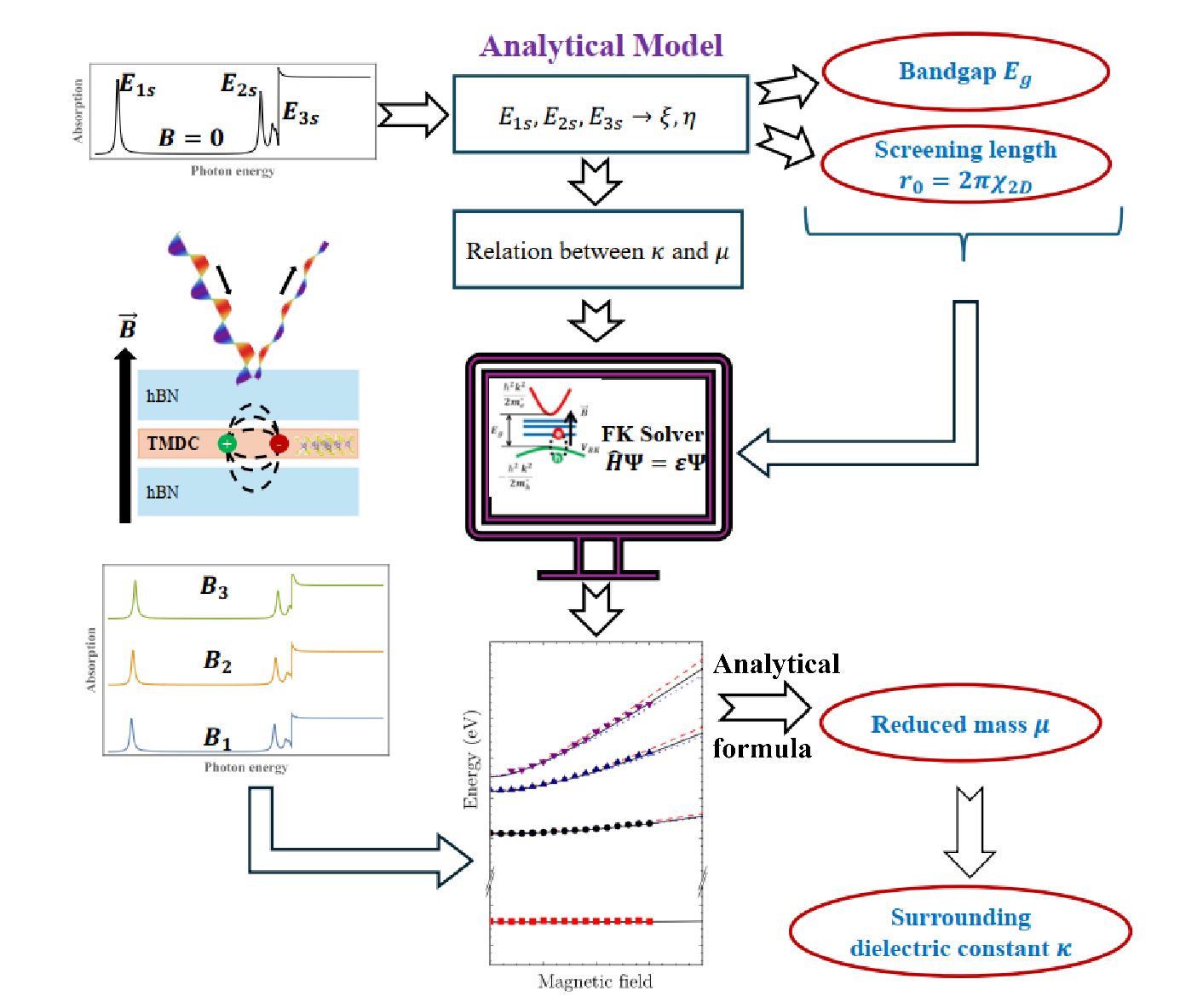}
\caption{Schematic flowchart of extracting the material parameters (the bandgap energy $E_g$, screening length $r_0$, exciton reduced mass $\mu$, and surrounding dielectric constant $\kappa$) from experimental exciton energies. Here, for the exciton reduced mass in the last step, we use the analytical formulation \eqref{eq17} and \eqref{eq18} for $\mu$ from the experimental data. }
\label{fig1}
\end{center}
\end{figure}

In the spirit of the availability of magnetoexciton spectroscopy for monolayer TMDCs, our group developed a method of probing material properties, $E_g$, $\mu$, $r_0$, and $\kappa$, from the magnetoexciton spectra \cite{nhat2019, nhat2023Apr, Nhat2023May}. Particularly, in Ref. \cite{Nhat2023May}, the exact numerical calculations of exciton energy spectra via the so-called Feranchuk-Komarov (FK) operator method \cite{giang1993, Hoangbook2015} allowed us to examine the sensitivity of magnetoexciton energies on the material parameters from which we established an efficient fitting scheme for probing material properties. Notwithstanding its accuracy, our numerical fitting scheme relied purely on extensive numerical calculations, preventing us from exploring the underlying physics of the method. Besides, due to the high computing resource consumption of the three-parameter fitting procedure, Ref.~\cite{Nhat2023May} provides only results for two TMDC monolayers: WSe$_2$ and WS$_2$.

Learning from the case of 3D semiconductors \cite{Miyata2015, Reimann1998}, we believe that an analytical model for exciton energy spectra can simplify our retrieval process and provide insight into the retrieved properties. Fortunately, an analytical expression for s-state exciton energy spectra at zero field was proposed within the regulated perturbation theory \cite{Dinh2025}. In the present study, this analytical formula enables us to derive a formulation for bandgap energy, screening length, and surrounding dielectric constant directly from the zero-field experimental exciton energies. These formulas also allow us to apply a one-parameter fitting procedure for the exciton reduced mass using magnetoexciton spectra, which significantly reduces the computer resources compared to the previous three-parameter fitting scheme. From other hand, with the achieved analytical formulas, the Schr{\"o}dinger equation for a magnetoexciton remains containing only one parameter, $\mu$. We can solve this equation to obtain an analytical expression for energies as functions of the exciton reduced mass $E(B,\mu)$. This one-variable function for each given magnetic intensity can be inversely transformed to obtain the exciton reduced mass as a function of exciton energies, which is a key factor in determining the exciton reduced mass from experimental magnetoexciton energies using an analytical method, without fitting. Figure \ref{fig1} illustrates our proposal to retrieve material parameters from experimental magnetoexciton energies.

 The rest of this paper is as follows. In Section \ref{Sec2}, we derive formulas for the bandgap energy $E_g$, screening length $r_0$, and surrounding dielectric constant $\kappa$. These formulas allow for the direct calculation of the bandgap energy $E_g$ and screening length $r_0$ from zero-field s-state exciton energies. Meanwhile, the formula for the surrounding dielectric constant $\kappa$ also contains the exciton reduced mass $\mu$. Section \ref{Sec3} develops an analytical method for the exciton reduced mass $\mu$ from magnetoexciton energies based on formulations derived in this section.
Section \ref{Sec4} presents the material parameters $E_g$, $r_0$, $\mu$, and $\kappa$ retrieved from the experimental exciton energies available for monolayer TMDCs using our method, and compares them with the available data.  Some discussions are also given. Other parameters, such as the diamagnetic constants and exciton radii calculated from the retrieved fundamental parameters, are also provided in this section.
Section \ref{Concl} presents our conclusions and outlook. 

\section{\label{Sec2}Analytical formulas for bandgap energy, screening length, and surrounding dielectric constant} 
Based on the Rytova-Keldysh potential describing the hole-electron interaction in monolayer TMDCs, Dinh {\it et al.} \cite{Dinh2025} proposed an analytical formula for the $s$-state exciton binding energies. For use in the present study, we rewrite it as   
\begin{eqnarray}\label{eq1}
\varepsilon_{ns} = -\eta \frac{P_n(\xi)}{\left( n - 0.5 + \xi \right)^2} \,\textrm {Ry},\quad n=1,2,3, \dots
\end{eqnarray}
where Ry $= 13605.69 $ meV is a Rydberg unit of energy; $P_n (\xi)={\textrm {exp}}(a_n \Delta)/(1+b_n \Delta)$ with $\Delta =0.225 /\xi -0.525$, $a_n=4.3/n -6.6/n^2+3.2/n^3$, and $b_n=5.3/n-7.6/n^2+4.0/n^3$. Here, the dimensionless parameter $\eta$ is the rescaling factor of Rydberg energy, while the dimensionless parameter $\xi$ captures the screening effect from dimensionality reduction. These parameters are defined as
\begin{equation}\label{eq2}
\eta={\frac{\mu}{\kappa^2}}, \quad \xi =0.479 \, \frac{r_0}{a_0}\frac{\mu}{\kappa^2},
\end{equation}
where $\mu=\mu^*/m_e$ is the dimensionless exciton reduced mass; $\kappa$ is the dielectric constant of the surrounding medium; $r_0$ is the screening length; $m_e$ is the electron mass; and $a_0= 0.0529$ nm is the Bohr radius. Here, the exciton reduced mass $\mu^*$ is calculated from the effective electron and hole masses ($m_e^*$, $m_h^*$) by the definition $\mu^*=m_e^* m_h^*/(m_e^*+m_h^*)$. We note that Formula~\eqref{eq1} works well in the range of $0.85> \xi >0.29$, which captures a broad class of monolayer TMDCs encapsulated by hBN slabs.

On the other hand, in the experimental photoabsorption spectrum, exciton peaks are located at 
\begin{equation}\label{eq3}
E_{ns}= \varepsilon_{ns}+ E_g,
\end{equation}
where $E_g$ is the bandgap energy of the TMDC monolayer, which needs to be retrieved.
Using Eqs. \eqref{eq1} and \eqref{eq3}, we can inversely define the bandgap  energy $E_g$ and parameters $\eta$, $\xi$ from the experimental exciton energies $E_{1s}$, $E_{2s}$, and $E_{3s}$ since there are three equations versus three unknowns, as shown by the following scheme:
\begin{equation}\label{eq4}
 {\textrm {Experimental}}\, (E_{1s}, E_{2s}, E_{3s}) \rightarrow   (\eta, \xi, E_g).
\end{equation}

\begin{figure}[htbp!]
\center
\includegraphics[width=1 \columnwidth]{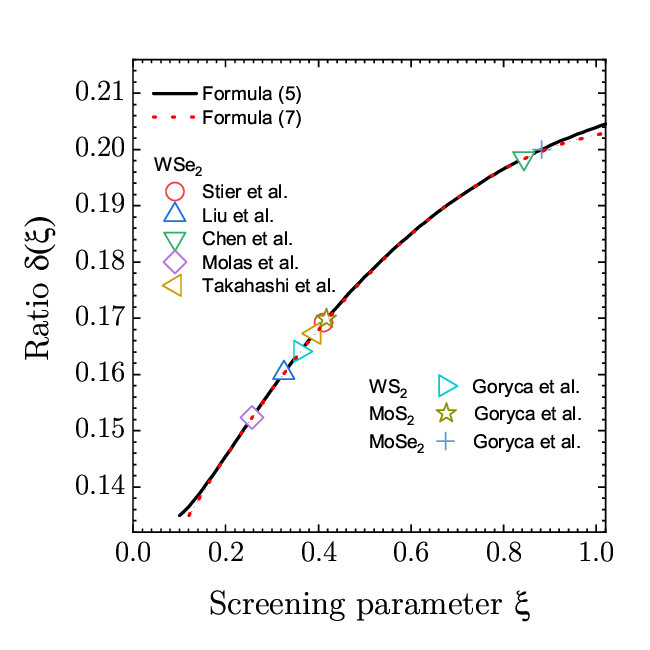}
\caption{Ratio $\delta(\xi)$ by Formula~\eqref{eq5} as a function of variable $\xi$ (black line) plotted with some experimental data       \cite{Stier2018,Liu2019,Chennano2019,Molas2019,takahashi2024, NAT2019} of $\delta_{\text{exp}} = (E_{2s} - E_{1s}) / (E_{3s} - E_{2s})$ marked by different symbols. Solution $\xi (\delta)$ \eqref{eq7} also plotted for comparison (red dots). }
\label{fig2}
\end{figure}

Indeed, from Eqs. \eqref{eq1} and \eqref{eq3}, we can find out that the following ratio 
\begin{eqnarray}\label{eq5}
\delta(\xi) &=&\frac{\varepsilon_{3s}-\varepsilon_{2s}}{\varepsilon_{2s}-\varepsilon_{1s}} =  \frac{E_{3s}-E_{2s}}{E_{2s}-E_{1s}} \nonumber\\
 &=&\frac{(1+2\xi)^2}{(5+2\xi)^2}\,\frac{(3+2\xi)^2P_3(\xi)-(5+2\xi)^2P_2(\xi)}{(1+2\xi)^2P_2(\xi)-(3+2\xi)^2P_1(\xi)}
\end{eqnarray}
is a function of only the screening parameter $\xi$ and does not depend on the Rydberg rescaling parameter $\eta$.
Therefore, the experimental data of $E_{1s}$, $E_{2s}$, and $E_{3s}$ for different materials such as WSe$_2$, WS$_2$, MoSe$_2$, and MoS$_2$ \cite{Stier2018, Liu2019, Chennano2019, takahashi2024, NAT2019,Molas2019} yield the experimental quantity $\delta_{exp}={(E_{3s}-E_{2s})}/(E_{2s}-E_{1s})$, from which we can determine the screening parameter $\xi$ by solving the following equation:
\begin{eqnarray}\label{eq6}
\delta(\xi) = \delta_{{exp}}\,.
\end{eqnarray}
For monolayer TMDCs on the hexagonal boron nitride (hBN) substrate, as studied in the most recent references \cite{Stier2018, Liu2019, Chennano2019, takahashi2024, NAT2019,Molas2019}, the quantity $\xi$ lies in the working range of Formula \eqref{eq1} for exciton binding energies.   Since the function $\delta(\xi)$ is single-valued and monotonically increases as $\xi$ increases in its range, as demonstrated in Fig.~\ref{fig2}, Eq.~\eqref{eq6} has a unique solution. This kind of equation is easy to solve numerically; however, we approximately solve it to obtain an analytical solution as:
\begin{eqnarray}\label{eq7}
\xi= 1.1219 -3.821\, \sqrt{0.2036-\delta_{exp}} \,.
\end{eqnarray}

After determining the screening parameter $\xi$ from Eq.~\eqref{eq6}, we can directly obtain the Rydberg rescaling parameter $\eta$  as well as the bandgap energy $E_g$ by solving Eq.~\eqref{eq3} for the first two states (n=1 and n=2). The results are
\begin{eqnarray}\label{eq8}
\eta = \frac{(1+2 \xi)^2 (3+2\xi)^2}{(3+2\xi)^2 P_1(\xi)- (1+2\xi)^2 P_2 (\xi)}\frac{E_{2s}-E_{1s}}{4\textrm{Ry}},
\end{eqnarray}
\begin{eqnarray}\label{eq9}
E_g = E_{1s} - \frac{(3+2\xi)^2 P_1 (\xi) \left(  E_{2s}- E_{1s}  \right)}{(1+2\xi)^2 P_2 (\xi) - (3+2\xi)^2 P_1 (\xi)}.
\end{eqnarray}
Using Eqs.~\eqref{eq8} and \eqref{eq9}, we can calculate $\eta$ and $E_g$ from the experimental exciton energies ($E_{1s}$, $E_{2s}$) and the value $\xi$, which is in turn determined from the experimental exciton energies $E_{1s}$, $E_{2s}$, and $E_{3s}$ by Eq.~\eqref{eq7}.

Knowing the parameters $\eta$ and $\xi$, we can directly achieve the screening length from Eqs.~\eqref{eq2} as
\begin{eqnarray}\label{eq10}
r_0 = 2.088\,\frac{\xi}{\eta}\, a_0\,.
\end{eqnarray}
Also, the surrounding dielectric constant $\kappa$ can be obtained from the first equation in Eqs.~\eqref{eq2}, which leads to the formula
\begin{eqnarray}\label{eq11}
\kappa = \sqrt{\frac{\mu}{\eta}}\,.
\end{eqnarray}
Equation \eqref{eq11} does not imply that the surrounding dielectric constant $\kappa$ is dependent on the exciton reduced mass $\mu$, but rather serves as a basis for our retrieval scheme only. 

For this Section, we conclude that the bandgap energy $E_g$ and screening length $r_0$ can be determined from the experimental exciton energies $E_{1s}$, $E_{2s}$, and $E_{3s}$ by Eqs.~\eqref{eq9} and \eqref{eq10}, independently of knowing the values of the surrounding dielectric constant $\kappa$ and the exciton reduced mass $\mu$. This conclusion also explains the numerical results of Ly {\it et al.} \cite{Nhat2023May}, where $r_0$ remains unchanged even when varying the values of $\kappa$ and $\mu$. Besides, the relationship \eqref{eq11} allows reducing one more material parameter, saying the surrounding dielectric constant $\kappa$, from the Schr{\"o}dinger equation. Consequently, these three formulas (Eqs.~\eqref{eq9}, \eqref{eq10}, and \eqref{eq11}) are essential because they remove the need to determine all structural parameters ($\mu$, $\kappa$, $r_0$, and $E_g$) simultaneously, leaving only the exciton reduced mass $\mu$.

\section{\label{Sec3} Exciton reduced mass from magnetoexciton energies}
To determine the exciton reduced mass $\mu$ from the available experimental magnetoexciton energies, Ref.~\cite{Nhat2023May} suggests a method based on fitting the calculated magnetoexciton energies with the experimental data while varying three parameters $\mu$, $r_0$, and $\kappa$. The analytical formulas \eqref{eq7}-\eqref{eq11} achieved in the present study allow reducing the fitting to a one-parameter problem. It saves computer resources significantly. However, more advanced, we suggest an analytical approach to calculate the exciton reduced mass $\mu$, based on the formula developed in this section for it as a function of magnetoexciton energies. 

For this purpose, first, we consider the Schr{\"o}dinger equation for an s-state exciton in monolayer TMDCs with the presence of a homogeneous magnetic field $B$ perpendicular to the TMDC plane as
\begin{eqnarray}\label{eq12}
\left\{ - \left( \frac{\partial^2}{\partial x^2}+\frac{\partial^2}{\partial y^2}\right)  
    +\frac{1}{8} B^2 (x^2+y^2) \right.\qquad\qquad\quad\nonumber\\
\left.  {- 2}
\int\limits_{0}^{+\infty} \frac{dq}{ \sqrt{1+\alpha^2 q^2} }\;
 \textrm{e}^{-qr}-\varepsilon\right\} \psi(x, y) = 0,\quad
\end{eqnarray}
where $\alpha=r_0/(\kappa a_0^*){=2.088\,\xi}$; $r=\sqrt{x^2+y^2}$; energy $\varepsilon$ and coordinates $x, y$ are given in the effective Rydberg unit $\text{Ry}^*=\mu m_e e^4/32\pi^2\kappa^2\epsilon_0^2\hbar^2$ and effective Bohr radius $a_0^*=4\pi\kappa\epsilon_0\hbar^2/\mu m_ee^2$, scaled by the exciton reduced mass $\mu$ and the surrounding dielectric constant $\kappa$; $B$ is dimensionless magnetic intensity measured in the unit of $B_0^*=(2\mu m_e/\hbar e) \text{Ry}^*$; $\hbar$ and $e$ are the reduced Planck constant and electron electric charge, respectively. 

Equation~\eqref{eq12} has been derived in Refs.~\cite{Nhat2023May,nhat2023Apr}, where the Laplace transform for the Rytova-Keldysh potential was used for describing the electron-hole interaction. In Ref.~\cite{nhat2023Apr}, the Schr{\"o}dinger equation was first established for the whole system, an electron and a hole interacting with each other through the Rytova-Keldysh potential. After that, the center of mass motion was separated by an exact procedure; thus, the Schr{\"o}dinger equation for the relative motion of an electron and a hole (an exciton with reduced mass $\mu$) was obtained. In the present paper, this equation is rewritten in a dimensionless form, where the material parameters $\mu$ and $\kappa$ consist only of scaled atomic units. The other parameter, $r_0$, is defined via $\alpha$, a dimensionless constant in the Laplace form of the Rytova-Keldysh potential.

For an analytical method, we need to solve the Schr{\"o}dinger equation \eqref{eq12} to obtain exciton energies as functions of the material parameters and the magnetic intensity, given as $\varepsilon_{ns} (B, \mu, \kappa, r_0)$. Using the relations described by Eqs.~\eqref{eq7}-\eqref{eq11}, we get the experimental exciton energy as a one-variable function $E_{ns} (\mu, B)$ for each value of the magnetic intensity $B$. Furthermore, we derive an inverse function $\mu (E_{ns}, B)$, which allows us to determine the exciton reduced mass $\mu$ from the measured exciton energies $E_{ns}(B)$. 

This kind of the Schr{\"o}dinger equation \eqref{eq12} can be resolved analytically by the regulated perturbation method, as developed in Refs.~\cite{Tram2013, Tram2016} for a two-dimensional exciton with the Coulomb potential for the electron-hole interaction in a magnetic field. However, in our case with the electron-hole interaction described by the Rytova-Keldysh potential, the problem is more complicated. We will find the magnetoexciton energy in the form
\begin{eqnarray}\label{eq13}
\varepsilon_{ns} (B)= \varepsilon_{ns} (0) + \Delta E_{ns}(B,\mu),
\end{eqnarray}
where the field-free exciton energy $\varepsilon_{ns} (0)$ is defined by Eq.~\eqref{eq1}. The magnetic part of energy  $\Delta E_{ns}(B,\mu)$ is obtained from the equation as a perturbative correction in the weak regime of the magnetic field ($B \ll 1$), given that the limit of magnetic intensity is less than 120 Tesla in most current experiments. 

Considering that perturbation theory, applied to the Schr\"{o}dinger equation \eqref{eq12}, provides the energy correction in the weak-field regime as a polynomial of magnetic intensity $B$, with two limits, $\sim B^2$ at the weak-field regime and $\sim B$ at the strong-field regime \cite{Tram2013, Tram2016}, we suggest the correction in the Padé approximant as
\begin{eqnarray}\label{eq14}
\Delta E_{ns}= \frac{\alpha_n B^2} {1+ \beta_n B} \,\text{Ry}^*\,,
\end{eqnarray}
where coefficients $\alpha_n$ and $\beta_n$ can be calculated by the second-order perturbation theory. We suggest an alternative approach by fitting the formula \eqref{eq14} with numerical exact solutions of the Schr{\"o}dinger equation \eqref{eq12}. Since the energy correction $\Delta E$ weakly depends on the material parameter $\xi$, the achieved coefficients $\alpha_n$ and $\beta_n$ are actually slightly material-dependent. We provide them in Table~\ref{tab1} for the case of $\xi=0.4135$ (corresponding to the experimental data of Stier {\it{et al.}} \cite{Stier2018}) and for $ns$ states,  $n=1, 2, 3, 4, 5$.

\begin{table}[H]
\caption{\label{tab1} Coefficients $\alpha_n$ and $\beta_n$ in the formula~\eqref{eq14} for $ns$ states ($n=1, 2, 3, 4, 5$) for the case of $\xi=0.4135$.}
\begin{ruledtabular}
\begin{tabular}{l c c c c c }
     $n$     &  1 &    2 &  3 & 4 & 5  \\
\hline \\[-5pt]
$\alpha_n$ & 0.5302  & 8.7698 &  45.2041 & 145.2032 & 217.8212 \\
$\beta_n$ & 0.4244  & 3.0909 &  10.1873 & 23.5106 & 32.9012 \\[2pt]

\end{tabular}
\end{ruledtabular}
\vspace{15pt}
\end{table}

In Eq.~\eqref{eq14}, the material parameters $\mu$ and $\kappa$ are hidden in the effective atomic units for energies and magnetic intensity: $\text{Ry}^*$ and $B_0^*$. To have this equation in an explicit form of the exciton reduced mass $\mu$, we utilize the standard atomic units  $\text{Ry}$ and $B_0$ instead of the effective units for the exciton energy and magnetic intensity, using the transformations:  $\text{Ry}^*=\mu m_e e^4/(32 \pi^2\epsilon_0^2\kappa^2\hbar^2)=(\mu/\kappa^2) \text{Ry} $ and $B_0^*= (2\mu/e\hbar) \text{Ry}^*=(\mu^2/\kappa^2) B_0$. As a consequence, we have 
\begin{eqnarray}\label{eq15}
\Delta E_{ns}(B,\mu)=\frac{1}{\mu}\frac{\alpha_{n}\,B^2}{\eta \mu + \beta_{n} B} \text{Ry}\,,
\end{eqnarray}
where the units of energy and magnetic intensity are now given by $\text{Ry}=m_e e^4/(32 \pi^2\epsilon_0^2 \hbar^2)=13.60569$ eV and $B_0=(2m_e/e \hbar) \text{Ry}=235,051.7$ Tesla, which are material-independent.  In Eq.~\eqref{eq15}, the value $\eta=\mu/\kappa^2$  is determined from the experimental field-free exciton energies by the formula \eqref{eq8}. 

Formula~\eqref{eq15} is highly accurate for a wide class of TMDC monolayers, including  WSe$_2$, WS$_2$, MoSe$_2$, and MoS$_2$ on hBN slabs. Compared with the exact numerical solutions, this formula gives an accuracy of less than 0.5 meV, so we can use it to retrieve the exciton reduced mass $\mu$ from experimental magnetoexciton energies. Indeed, given the formula for the exciton energy $E_{ns}(B)= \varepsilon_{ns}(0) + \Delta E_{ns} (B,\mu) +E_g$, the quantity $\delta_B$, defined by the following equation
\begin{eqnarray}\label{eq16}
\delta_B&=&\frac{\Delta E_{ns}(B_1,\mu)-\Delta E_{ns}(B_2,\mu)}{\alpha_n (B_1-B_2)}  \nonumber\\
&=&\frac{E_{ns}(B_1)-E_{ns}(B_2)} {\alpha_n(B_1-B_2)}=\delta_B^{exp},
\end{eqnarray}
can be obtained from any two points of experimental data $E(B_1)$ and $E(B_2)$ at magnetic field values $B_1$ and $B_2$. Using the formula~\eqref{eq15}, we lead Eq.~\eqref{eq16} to an equation for the exciton reduced mass $\mu$ as
\begin{eqnarray}\label{eq17}
\mu^3+a \mu^2 + b\mu +c=0
\end{eqnarray}
with 
\begin{eqnarray}\label{eq18}
a&=& \frac{\beta_n}{\eta} (B_1+B_2),\nonumber\\
b&=& \frac{{\beta_n}^2}{\eta^2} B_1 B_2- \frac{1}{\eta\,\delta_B^{exp} } (B_1+B_2),\nonumber\\
c&=& - \frac{\beta_n}{\eta^2\delta_B^{exp} } B_1 B_2 .
\end{eqnarray}
Equation~\eqref{eq18} is easy to solve for every two picked data points, and the final result is an average of all solutions.

We can explain the physical meaning of the coefficients $\alpha_n$, which are associated with the diamagnetic constants, defined by the equation:
\begin{eqnarray}\label{eq19}
\sigma_{ns}&=&\frac{1}{2} \mathop{\lim}\limits_{B \to 0} \frac{\partial^2 \Delta E_{ns}}{{\partial B}^2}
=\frac{\alpha_{n} \kappa^2}{\mu^3}\times \frac{{\mu_B}^2}{\text{Ry}},
\end{eqnarray}
where $\mu_B=e\hbar/2m_e=5.788381 \times 10^{-2}$ meV/T is the Bohr magneton; and ${\mu_B}^2/{\text{Ry}}=0.24626 \times 10^{-3}\, \mu{\text{eV}}/{\text{T}}$. Equation~\eqref{eq19} can be used as an analytical formula for diamagnetic constants, which agrees excellently with the numerical calculation obtained using the traditional formulation $\sigma_{ns}=\frac{e^2}{8\mu m_e} {r_{ns} }^2$, where the exciton radii are calculated by equation $r_{ns}=\sqrt{\langle \psi_{ns}| r^2 |\psi_{ns}\rangle}$, with $\psi_{ns}$ representing field-free wave functions. Using Eq.~\eqref{eq19}, we also arrive to a formula for exciton radii as
\begin{eqnarray}\label{eq20}
r_{ns}&=& \frac{2\kappa}{\mu} \sqrt{\alpha_n}\,a_0,
\end{eqnarray}
where $a_0=0.0529$ nm is the Bohr radius.

\section{\label{Sec4} Fundamental material parameters of monolayer TMDCs from the available experimental magnetoexciton energies} 
 From the above-formulated analytical expressions, we derive a practical scheme, as shown in Fig.~\ref{fig1}, for retrieving material parameters from experimental magnetoexciton energies.

Applying this retrieval scheme to the experimental data of Refs.~\cite{Stier2018,Liu2019,Chennano2019,Molas2019,takahashi2024}, we retrieve the material parameters for monolayer WSe$_2$. We provide them in Table~\ref{tab2} and compare them with those presented in the references. More data retrieved in the same manner from Ref.~\cite{NAT2019} for other TMDC monolayers, such as WS$_2$, MoS$_2$, and MoSe$_2$, are also provided in Table~\ref{tab3} and compared with others.
Knowing the material parameters, we also calculate other essential properties, such as the diamagnetic constants and exciton radii, using equations \eqref{eq19} and \eqref{eq20}. Table~\ref{tab4} presents the results obtained in our study, which are compared with experimental and theoretical data from references.

The first remark from Table~\ref{tab2} is that although various experiments were performed for the same material, monolayer WSe$_2$ encapsulated by hBN slabs, the surrounding dielectric constants $\kappa$  from these experiments differ slightly due to variations in the experimental setups. However, the exciton reduced mass $\mu$ is weakly dependent on the surrounding dielectric constant $\kappa$, as shown in Table~\ref{tab2}. For various experiments, the value is almost $0.19 \,m_e$, differing from the value, $0.20 \,m_e$, used in these experiments for theoretical explanations. 
With the exciton reduced mass known, our method is also applicable to photoluminescence experiments without a magnetic field, as shown in Table~\ref{tab2} for Ref.~\cite{takahashi2024}. We also note that the exciton reduced mass value in monolayer WSe$_2$, as calculated in the present study from the experimental data, agrees better with the theoretical calculations by density functional theory (DFT) \cite{berkelbach2013, kylanpaa2015}.

The second remark is about the surrounding dielectric constant, which indeed is independent of the materials. For various experiments \cite{Stier2018, Liu2019, Molas2019, takahashi2024, NAT2019} with different TMDC monolayers (WSe$_2$, WS$_2$, MoSe$_2$, and MoS$_2$), the results are nearly identical (except in Chen {\it et al.} \cite{Chennano2019}), with values approximately equal to the bulk h-BN dielectric constant of 4.5, as referenced in \cite{HePRL2014}.

We also note the slight discrepancy in Tables~\ref{tab2} and \ref{tab3} between the bandgap energies calculated by our method and those in the references.   By applying the bandgap energies we calculated, the experimental binding energies for the $1s$ state (in parentheses) coincide with the prediction of our theory. The consistency of our predicted energy transitions $\Delta E_{1s-2s}$ and $\Delta E_{2s-3s}$ to the experimental data is much better than the theoretical modeling by the Keldysh potential in Refs.~\cite{Stier2018, Liu2019, Chennano2019} or by the Kratzer potential in Ref.~\cite{Molas2019} for the electron-hole interaction. Our results are also consistent with our previous study \cite{Nhat2023May} for monolayers WSe$_2$ and WS$_2$.  

\onecolumngrid

\begin{table}
\caption{\label{tab2} Fundamental optoelectronic material parameters (bandgap energy $E_g$, screening length $r_0=2\pi \chi_{\text {2D}}$ with 2D static polarizability $\chi_{\text {2D}}$, exciton reduced mass $\mu$, surrounding dielectric constant $\kappa$) for monolayer WSe$_2$ retrieved from magnetoexciton energies of different experiments \cite{Stier2018,Liu2019,Chennano2019,Molas2019,takahashi2024} using analytical expressions \eqref{eq9}, \eqref{eq10}, \eqref{eq11}, and \eqref{eq17} and compared with those given in these references.}
\begin{ruledtabular}
\begin{tabular}{l c c c c c c c c}
   WSe$_2$   &  {$\xi$} &$r_0$ &     $\mu$   & $\kappa$&  $E_g$ &$E_b$& $\Delta E_{1s-2s}$& $\Delta E_{2s-3s} $\\
   &&   nm &     $(m_e)$   & &   meV &meV  &eV& meV\\
\hline \\[-5pt]
Stier {\it et al.}  -- exp.  &--& --& -- &  -- &-- &(169)&130 &22\\
 -- theory  \cite{Stier2018} &--& 4.50& 0.20 &  4.50 &1.890 &167&124 &21.3\\
Present work's theory  & {0.4135}&4.35&0.19 &  4.25& 1.892&168.8& 130.0 &22.0\\[2pt]
\hline \\[-5pt]
Liu {\it et al.}  -- exp.  &--& -- & -- &  --  & --&(167)& 131& 21\\
 -- theory  \cite{Liu2019}&--& 5.00&0.20 &  3.97  & 1.884&172& 128.3 &23.3\\
Present work's theory &{0.3268}& 3.89&0.185 &  4.47  & 1.879  &167.4& 131.0&21.0\\[2pt]
\hline \\[-5pt]
Chen {\it et al.}  -- exp.  &--& --&--&  -- & --&(180)&131 &26 \\ 
 -- theory  \cite{Chennano2019} &--& 4.51&  0.22 & 4.50&1.900&170&128 &23 \\
Present work's theory  &{0.8483}& 5.52& 0.19  & 3.35& 1.907&180.0& 131.0&26.0 \\[2pt]
\hline \\[-5pt]
Molas {\it et al.} -- exp. &--& --& --  & -- & --&(167)&129 &21.5\\
 -- theory  \cite{Molas2019} &--& 4.50& 0.20  & 4.50 & 1.873&167&128.0 &21.2 \\
Present work &{0.3876}& 4.26& 0.19 & 4.35 & 1.873 &166.7&129.0&21.5\\[2pt]
\hline \\[-5pt]
Takahashi {\it et al.} -- exp.  &--& -- &-- &  -- &--& (167)&129.1 &21.6\\
-- theory  \cite{takahashi2024} &--& 3.45 &0.20 &  5.01 &1.862&164&131 &19\\
Present work  &{0.3940}& 4.29 & 0.19* &  4.33 &1.865&167.0& 129.1&21.6\\[2pt]

\end{tabular}
\end{ruledtabular}
\end{table}

\begin{table} [thbp]
\caption{\label{tab3} Fundamental optoelectronic material parameters (bandgap energy $E_g$, screening length $r_0=2\pi \chi_{\text {2D}}$ with 2D static polarizability $\chi_{\text {2D}}$, exciton reduced mass $\mu$, surrounding dielectric constant $\kappa$) for monolayer WS$_2$, MoSe$_2$, and MoS$_2$ retrieved from magnetoexciton energies of the experiment \cite{NAT2019} using analytical expressions \eqref{eq9}, \eqref{eq10}, \eqref{eq11}, and \eqref{eq17} and compared with those given in this reference.}
\begin{ruledtabular}
\begin{tabular}{l c c c c c c c c}
      &{$\xi$}&  $r_0$ &     $\mu$   & $\kappa$&  $E_g$ &$E_b$& $\Delta E_{1s-2s}$& $\Delta E_{2s-3s} $\\
   &&   nm &     $(m_e)$   & &   eV &meV  &meV& meV\\
\hline \\[-5pt]
WS$_2$, Goryca {\it et al.}  -- exp. & --& --& -- &  -- &-- &(181)&141 &23\\
 -- theory  \cite{NAT2019} &--& 3.40& 0.175 &  4.35 &2.238 &180&-- &--\\
Present work's theory   &{0.3531}&3.74&0.175 &  4.10& 2.239&181.0& 141.0  &23.0\\[2pt]
\hline \\[-5pt]
MoSe$_2$, Goryca {\it et al.} -- exp.  &--& --& -- &  -- &-- &(232)&171 &33\\
-- theory  \cite{NAT2019}  &--& 3.90& 0.35 &  4.40 &1.874 &231&-- &--\\
Present work's theory   &{0.7045}&4.02&0.345 &  4.22& 1.875&231.7& 171.2  &32.8 \\[2pt]
\hline \\[-5pt]
MoS$_2$, Goryca {\it et al.}  -- exp.  &--& --& -- &  -- &-- &(221)&170 &29\\
 -- theory  \cite{NAT2019}&-- & 3.40& 0.27 &  4.40 &2.161 &222&-- &--\\
Present work's theory   &{0.4276}&3.37&0.275 &  4.43& 2.160&221.3& 170.0  &29.0 \\[4pt]

\end{tabular}
\end{ruledtabular}
\vspace{15pt}
\end{table}

\twocolumngrid

We emphasize that the diamagnetic constants and corresponding exciton radii can be calculated analytically by Eqs.~\eqref{eq19} and \eqref{eq20}, which are very close to the ones calculated numerically, as shown in Table~\ref{tab4}. For the 1s state, our theoretical results as well as the theoretical modelling of others agree well with the experimental data, except for the experiment of Ref.~\cite{Chennano2019}, where, however, our theoretical calculations are better than those in this reference. For the excited states, the experimental data are subject to considerable uncertainty due to the lack of data near the zero-point magnetic field $B=0$. In these cases, our calculations are relatively consistent with the experimental data and more so than those in the references.

\begin{figure}[H]
\center
\includegraphics[width=1. \columnwidth]{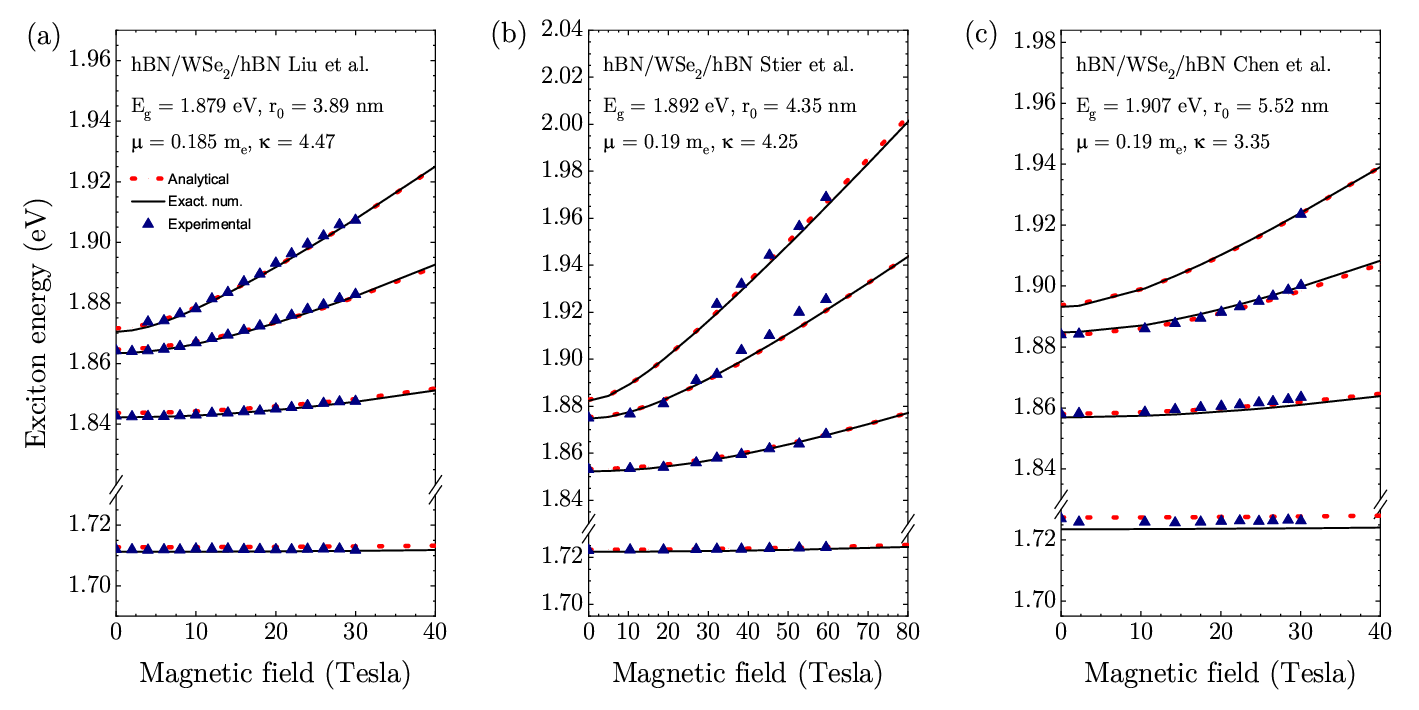}
\caption{Magnetoexciton energies in monolayer WSe$_2$ encapsulated by hBN slabs, calculated by the highly accurate numerical method (black lines) and by the analytical method (formulas \eqref{eq1} and \eqref{eq15}) (red lines) compared with the experimental data (blue triangles) of Refs.~\cite{Stier2018} (a), \cite{Liu2019} (b), and \cite{Chennano2019} (c). Material parameters used in the calculations are retrieved from the experimental data using the formulas \eqref{eq9} (for the band gap energy $E_g$), \eqref{eq10} (for the screening length $r_0$), \eqref{eq11} (for the dielectric constant $\kappa$), and \eqref{eq17} (for the exciton reduced mass $\mu$).}
\label{fig3}
\end{figure}

\begin{figure}[H]
\center
\includegraphics[width=1. \columnwidth]{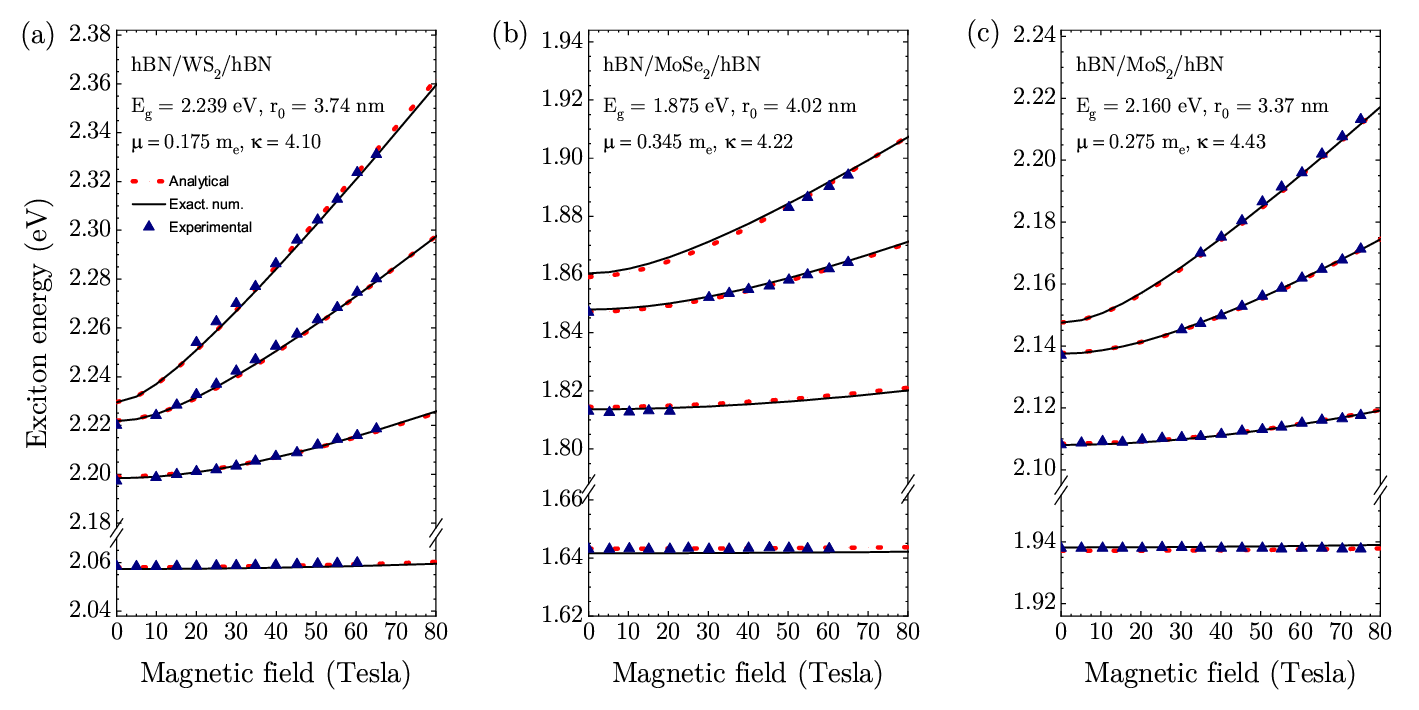}
\caption{Magnetoexciton energies of monolayer WS$_2$ (a), MoSe$_2$ (b), and MoS$_2$ (c), calculated numerically (black lines) and analytically (red lines) with the retrieved material parameters and compared with the experimental data (blue triangles) in Ref.~\cite{NAT2019}. Material parameters used in the calculations are retrieved from the experimental data using the formulas \eqref{eq9} (for the band gap energy $E_g$), \eqref{eq10} (for the screening length $r_0$), \eqref{eq11} (for the dielectric constant $\kappa$), and \eqref{eq17} (for the exciton reduced mass $\mu$).}
\label{fig4}
\end{figure}

Finally, we validate the retrieved material parameters in our study by plotting the magnetoexciton energies calculated using these parameters and comparing them with the experimental data. The results are provided in Figs.~\ref{fig3} and \ref{fig4}. Figure~\ref{fig3} presents our calculations compared with experimental magnetoexciton energies in monolayer WSe$_2$ \cite{Stier2018, Liu2019, Chennano2019}. Figure~\ref{fig4} demonstrates the comparison for magnetoexciton energies in monolayers WS$_2$, MoSe$_2$, and MoS$_2$ with the experiments of Ref.~\cite{NAT2019}. In these figures, theoretical data are calculated by two different methods: (i) by numerically solving the Schr\"{o}dinger equation \eqref{eq12} and (ii) by analytical formulas \eqref{eq1} and \eqref{eq15}. The figures obtained from the two methods are almost identical, demonstrating the high accuracy of our analytical solutions. The only discrepancy between the two methods is in the exciton energies at zero field, where the analytical formula \eqref{eq1} yields slight shifts (less than 0.5 meV) in energy compared to the exact numerical values. 

\onecolumngrid

\begin{table}[thbp]
\caption{\label{tab4} Diamagnetic constants $\sigma_{ns}$ (in units of $\mu$eV/T$^2$) and exciton radii $r_{ns}$ (in units of nm), calculated using the material parameters of TMDC monolayers WSe$_2$, WS$_2$, MoSe$_2$, and MoS$_2$ in the present study and compared with the experimental and theoretical data in Refs.~\cite{Stier2018,Liu2019,Chennano2019,NAT2019}.}
\begin{ruledtabular}
\begin{tabular}{l c c c c c c c c c c}
      & $\mu$ &  $\kappa$ &  $\sigma_{1s}$ & $r_{1s}$ &  $\sigma_{2s}$ & $r_{2s}$ & $\sigma_{3s}$& $r_{3s}$ & $\sigma_{4s}$ & $r_{4s}$\\
   & &   &     &    &   &  &&    & &  \\
\hline \\[-5pt]
WSe$_2$ -- exp.       &  --    & --    & 0.31 $\pm$ 0.02   &  1.7 $\pm$ 0.1     & 4.6 $\pm$ 0.2  &  6.6 $\pm$ 0.4 & 22$\pm$ 2 & 14.3$\pm$1.5 &--&--\\
 -- theory  (Stier \cite{Stier2018}) & 0.20 & 4.50 & 0.30                     &  1.67                    & 5.12                & 6.96 &26.86 &15.8&--&--\\
Our theory -- anal.   & 0.19 & 4.25 & 0.344 &  1.723& 5.687 &7.008 & 29.315 & 15.911&94.164&28.517\\
-- numer. method   & 0.19 & 4.25 & 0.329 &  1.686& 5.633 &6.977 & 29.393  & 15.938&94.879&28.635\\[2pt]
\hline \\[-5pt]
WSe$_2$ -- exp. & --& --& 0.24 $\pm$ 0.1 & 1.6 $\pm$ 0.4  & 6.4 $\pm$ 0.2 & 8.24 $\pm$ 0.13&27.3$\pm$ 1.3 &17.0 $\pm$ 0.4&73.7 $\pm$ 3.0& 27.8$\pm$ 0.7\\
 -- theory  (Liu \cite{Liu2019}) &0.20& 3.97& 0.31 &  1.68 &4.86 &6.66&24.2 &14.86&76.3&26.37\\
Our theory -- anal.   & 0.185 & 4.47&0.341 & 1.694& 6.124&7.176& 32.911  &16.636 &110.476&30.480\\
-- numer. method   & 0.185 & 4.47&0.336 & 1.682& 6.118&7.175& 32.905  &16.640 &107.920&30.135\\[2pt]
\hline \\[-5pt]
WSe$_2$ -- exp. & --& --& 0.5 &  2.2 & 5.8 & 7.6&17.6 &13.3&--&--\\
 -- theory (Chen \cite{Chennano2019}) &0.22& 4.50& 0.25 &  1.60 &4.18 &6.50&21.6 &14.7&--&--\\
Our theory -- anal.   & 0.19&3.35&0.348&  1.733& 4.916&6.516&23.128 &14.133 &70.390&24.656\\
-- numerical   & 0.19&3.35          &0.342& 1.718& 4.895&6.504& 23.096 &14.128 &70.389&24.664\\[2pt]
\hline \\[-5pt]
WS$_2$ -- exp. & --& --& -- &  -- &-- &--&-- &--&--&--\\
 -- theory (Goryca  \cite{NAT2019}) &0.175& 4.35& 0.40 &  1.8 &-- &--&-- &--&--&--\\
Our theory -- anal.   & 0.175& 4.10&0.369 &  1.712& 6.257&7.055& 33.407 &16.301 &109.296&29.486\\
-- numerical   & 0.175& 4.10&0.352 &  1.673& 6.272&7.066& 33.407 &16.307 &109.011&29.457\\[2pt]
\hline \\[-5pt]
MoSe$_2$ -- exp.  &--& --& -- &  -- &-- &--&-- &--&--&--\\
-- theory  (Goryca \cite{NAT2019}) &0.35& 4.40& 0.07 &  1.1 &-- &--&-- &--&--&--\\
Our theory -- anal.   &0.345&4.22&0.080 &  1.117& 1.180&4.302& 5.714  &9.466 &17.731&16.675\\
-- numerical   &0.345&4.22&0.079 &  1.114& 1.183&4.309& 5.729  &9.482 &17.732&16.681\\[2pt]
\hline \\[-5pt]
MoS$_2$ -- exp.  &--& --& -- &  -- &-- &--&-- &--&--&--\\
 -- theory  (Goryca \cite{NAT2019})&0.27 & 4.40& 0.12 &  1.2 &-- &--&-- &--&--&--\\
Our theory -- anal.   &0.275&4.43&0.119 &  1.222& 2.050&5.062& 10.678  &11.553 &34.427&20.745\\
-- numerical   &0.275&4.43&0.120 &  1.227& 2.051&5.065& 10.680  &11.558 &34.429&20.752\\[4pt]

\end{tabular}
\end{ruledtabular}
\vspace{15pt}
\end{table}

\twocolumngrid

\onecolumngrid

\twocolumngrid

\onecolumngrid

\twocolumngrid

There is a relatively significant shift of approximately 4.0  meV between numerical and analytical calculations of the bound state in the {Chen \it {et al.}}  experiment (Fig.~\ref{fig3} (c)). However, the extrapolation from the magnetoexciton energies yields an estimated value of 1,724 meV,  compared to the experimental value of 1,727 meV for the $1s$-state energy, which leads to a discrepancy in calculating the material parameter. Indeed, using the extrapolated value of the bound state energy, we obtain  $\xi=0.7481$, $\kappa=3.47$, $r_0=5.23$ nm, and $\mu=0.19$. The bandgap energy is now calculated as $E_g=1906.3$ meV, which leads to a bound energy of 182.32 meV.  By incorporating the new parameters, our analytical and numerical calculations for the ground-state energies are much more consistent.

\begin{figure}
\includegraphics[width=0.95 \columnwidth]{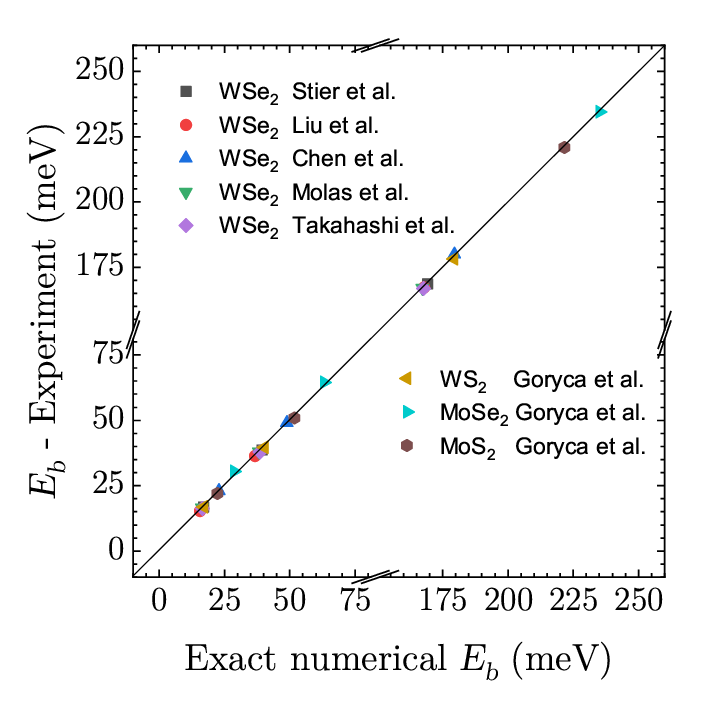}
\caption{Exciton binding energies calculated by numerical method with the retrieved material parameters for monolayer TMDCs in Tables~\ref{tab2} and \ref{tab3} and compared with the experimental in Refs.~\cite{Stier2018,Liu2019,Chennano2019,Molas2019,takahashi2024, NAT2019}. }
\label{fig5}
\end{figure}

Besides, we also confirm the validity of the retrieved material parameters for the free-field case when $B=0$. Putting the complete parameters ($\mu$, $\kappa$, and $r_0$) into the Schr{\"o}dinger equation \eqref{eq12} and solving it with $B=0$ using the method given in Ref.~\cite{Nhat2023May}, we obtain exact numerical exciton energies and compare them with the experimental ones used to retrieve the material parameters. We show the consistency between the theory and experiments in Fig.~\ref{fig5} for monolayer TMDCs used in Tables~\ref{tab2} and \ref{tab3} \cite{Stier2018,Liu2019,Chennano2019,Molas2019,takahashi2024, NAT2019}. 

We note the boundary of our theory, which is related to the working range  $0.85> \xi >0.29$ of the formula \eqref{eq1} for the $s$-state exciton energies. For all experiments involving monolayer TMDCs in an hBN surrounding environment, including those considered in the present study, the value $\xi$ falls in this range. However, for the experiment of Chernikov {\it{et al.}} \cite{chernikov2014}, where the monolayer WS$_2$ lies on a thin SiO$_2$ layer, the screening parameter $\xi\sim 5.0$ falls outside the range. As shown in Ref.~\cite{Dinh2025}, the working range ensures good precision with a relative error of less than $5 \%$. However, for values far outside the range, the calculation error may be significant. We will further develop the analytical formula for exciton energies to ensure its working range encompasses both free-standing and other surrounding media, including those of current interest \cite{Mhenni2025}.  

\section{\label{Concl}Conclusion} We have successfully developed an analytical approach for retrieving fundamental material parameters, such as bandgap energy, screening length, exciton reduced mass, and surrounding dielectric constant, from experimental magnetoexciton energies in monolayer transition metal dichalcogenides (TMDCs). Unlike previous fitting methods that required three parameters, our approach allows for the independent determination of the bandgap energy and screening length from the field-free $s$-state exciton energies. This results in a one-parameter problem, with the exciton reduced mass being the only parameter that needs to be defined. Notably, for the first time, the bandgap energy and screening length can be computed directly from exciton energies using straightforward analytical formulas. 

Furthermore, using perturbation theory combined with the Padé approximant, we have derived an analytical expression for magnetoexciton energies as functions of the exciton reduced mass.  This allows for the extraction of the exciton reduced mass analytically from experimental data on magnetoexciton energies. The surrounding dielectric constant is then also determined through an analytical formula. Besides, we have also developed analytical expressions for diamagnetic coefficients and exciton radii using the extracted material data. This analytical theory for magnetoexciton energies is significant and valuable for analyzing experimental data. 

By applying our method to the currently available experimental data for exciton energies in monolayer TMDCs encapsulated by hBN, we have retrieved fundamental material properties that complement and correct the existing data. The computation process is straightforward and user-friendly, making it suitable for other experiments aimed at retrieving fundamental material properties.

\section*{Acknowledgments} This research is funded by the Vietnam National Foundation for Science and Technology Development (NAFOSTED) under Grant No. 103.01-2023.138 and carried out by the high-performance cluster at Ho Chi Minh City University of Education, Vietnam.

Contribution: V.-H.L., D.-N.Ly, and D.-N.Le conceptualized the work and performed analytical formulation; D.-N.Ly, D.-K.D.L., and V.-H.L. carried out numerical calculations; V.-H.L., D.-N.Ly, D.-N.Le, and D.-K.D.L analyzed data, V.-H.L., D.-N.Ly, and D.-N.Le wrote and edited the manuscript.


\bibliography{ref}

\end{document}